\documentclass[conference]{IEEEtran}
\IEEEoverridecommandlockouts

\usepackage{cite}
\usepackage{amsmath,amssymb,amsfonts}
\usepackage{algorithmic}
\usepackage{graphicx}
\usepackage{textcomp}
\usepackage{xcolor}
\usepackage{algorithm}
\usepackage{algorithmic}
\usepackage{tabularx,array}
\usepackage{booktabs}
\usepackage{multirow} 
\usepackage{newfloat}
\usepackage{listings}
\usepackage[colorlinks=true, urlcolor=blue, citecolor=blue, linkcolor=blue]{hyperref}
\def\BibTeX{{\rm B\kern-.05em{\sc i\kern-.025em b}\kern-.08em
    T\kern-.1667em\lower.7ex\hbox{E}\kern-.125emX}}
\begin{document}

\title{Decoding Translation-Related Functional Sequences in 5'UTRs Using Interpretable Deep Learning Models\\

\thanks{\textsuperscript{*}These authors contributed equally.\par
\textsuperscript{\dag}Corresponding author.}

}
\author{
\IEEEauthorblockN{%
Yuxi Lin\textsuperscript{*},\;
Yaxue Fang\textsuperscript{*},\;
Zehong Zhang\textsuperscript{*},\;
Zhouwu Liu,\;
Siyun Zhong,\;\\
Zhongfang Wang\textsuperscript{\dag},\;
Fulong Yu\textsuperscript{\dag}}
\IEEEauthorblockA{%
Guangzhou National Laboratory, Guangzhou, Guangdong, China\\
\{lin\_yuxi, fang\_yaxue, liu\_zhouwu, zhong\_siyun, wang\_zhongfang, yu\_fulong\}@gzlab.ac.cn\\
2025390011@gzhmu.edu.cn}}

\maketitle

\begin{abstract}
Understanding the regulatory role of 5' untranslated regions (5'UTRs) in mRNA translation is critical for controlling protein expression and designing effective therapeutic mRNAs. While recent deep learning models have shown promise in predicting translational efficiency from 5'UTRs sequences, most are constrained by fixed input lengths and limited interpretability. We introduce UTR-STCNet, a Transformer-based architecture for flexible and biologically grounded modeling of variable-length 5'UTRs. UTR-STCNet integrates a Saliency-Aware Token Clustering (SATC) module that iteratively aggregates nucleotide tokens into multi-scale, semantically meaningful units based on saliency scores. A Saliency-Guided Transformer (SGT) block then captures both local and distal regulatory dependencies using a lightweight attention mechanism. This combined architecture achieves efficient and interpretable modeling without input truncation or increased computational cost. Evaluated across three benchmark datasets, UTR-STCNet consistently outperforms state-of-the-art baselines in predicting mean ribosome load (MRL), a key proxy for translational efficiency. Moreover, the model recovers known functional elements such as upstream AUGs and Kozak motifs, highlighting its potential for mechanistic insight into translation regulation.The code is available on \href{https://github.com/Yu-Lab-Genomics/UTR-STCNet}{GitHub}\footnote{\url{https://github.com/Yu-Lab-Genomics/UTR-STCNet}}.

\end{abstract}

\begin{IEEEkeywords}
5' untranslated regions, translational efficiency, interpretable deep learning
\end{IEEEkeywords}

\section{Introduction}

Messenger RNA (mRNA) therapeutics have emerged as a powerful and promising platform for the precise, tunable, and designable control of protein expression, with broad applications in vaccine development, cancer immunotherapy, and the treatment of genetic disorders \cite{qin2022mrna}.
The ultimate efficacy of mRNA-based therapeutics is primarily determined by their translation efficiency, which dictates the amount and functionality of proteins synthesized from engineered mRNA constructs.
Among the multiple factors influencing translation, the 5' untranslated region (5'UTRs) serves as a key regulatory element that plays a central role in translation initiation \cite{araujo2012before}.
Its sequence features, such as upstream AUGs (uAUGs) \cite{wang20045} and Kozak consensus motifs \cite{li2017nucleotides}, can significantly affect ribosomal scanning dynamics and start-site selection, thereby modulating the overall efficiency of translation.
Therefore, a systematic understanding of the regulatory code encoded within 5'UTRs is essential for the rational design and optimization of therapeutic mRNA constructs \cite{cao2021high}.


Recent advances in high-throughput experimental techniques, including ribosome profiling \cite{brar2015ribosome} and systematic mutagenesis \cite{kang2004systematic}, have produced large-scale datasets that link 5' untranslated region (5'UTRs) sequences to translational outcomes.
These high-quality resources have enabled researchers to apply deep learning models to systematically explore the relationship between 5'UTRs sequences and translation efficiency \cite{zhang2017analysis}, with mean ribosome load (MRL) frequently used as a key quantitative indicator of translation efficiency.
The development of these models has not only deepened the quantitative understanding of translation regulation but also laid the groundwork for subsequent computational studies in mRNA design.
Despite encouraging progress, current models still face two critical limitations \cite{sample2019human}.



First, most existing architectures are trained on fixed-length input sequences and therefore lack the ability to generalize to native 5' untranslated regions (5'UTRs) of varying lengths. To accommodate such constraints, sequences are often truncated or padded to match the input window, which may mask or distort key regulatory elements. This preprocessing can lead to the loss of critical contextual information, especially in sequences containing long regulatory segments or multiple functional sites. As a result, models often struggle to capture length-dependent features present in natural 5'UTRs, reducing both prediction stability and generalization performance.
Second, current models offer limited interpretability. Although some incorporate attention mechanisms or structural priors, they typically fail to assign functional effects to specific sequence motifs, such as upstream AUGs (uAUGs) or Kozak elements \cite{kozak1987analysis}, thereby restricting their usefulness in mechanistic inference and variant prioritization. This limitation prevents models from providing biologically grounded explanations, leaving researchers reliant on overall prediction metrics rather than understanding which sequence features truly drive translation efficiency differences.

In addition, current models often struggle to scale effectively in multi-task learning scenarios or to operate efficiently in resource-constrained environments, limiting their applicability in complex biological systems and real-world deployment contexts. In practical applications, such computational bottlenecks not only increase training and inference costs but also constrain the operational efficiency of models in high-throughput tasks.


To address these challenges, we present UTR-STCNet, a Transformer-based deep learning framework for predicting translational efficiency from 5' untranslated region (5'UTRs) sequences.
UTR-STCNet introduces two key innovations:
a Saliency-Aware Token Clustering (SATC) module that groups and filters sequence tokens based on regulatory relevance, and a Saliency-Guided Transformer (SGT) block that models motif interactions across both local and long-range contexts using sparse attention.
This design enables the model to accurately capture translation-relevant signals from variable-length 5'UTRs while maintaining high biological interpretability and computational efficiency.
Moreover, UTR-STCNet balances performance and scalability in its architectural design, exhibiting stable predictive performance across different data scales and task settings. The framework not only achieves significant improvements in prediction accuracy but also provides a new analytical perspective for understanding 5'UTRs regulatory mechanisms.
Our contributions are summarized as follows:
\begin{itemize}
  \item We propose UTR-STCNet, a deep learning framework for predicting mean ribosome load from 5'UTRs sequences, with strong generalization across datasets and biological contexts.

  \item The model supports variable-length inputs, scales to multi-task settings, and maintains a lightweight architec- ture suitable for large-scale or resource-constrained de- ployment.
  
  \item UTR-STCNet offers intrinsic interpretability by explic- itly identifying regulatory motifs such as uAUGs and Kozak sequences.
  
  \item UTR-STCNet consistently outperforms existing baselines across tasks involving multiple species and cell lines, achieving state-of-the-art performance in translation prediction.
\end{itemize}

\section{Related Work}

\subsection{Self-Supervised Representation Learning for RNA Sequences}
A growing body of work has focused on learning contextual representations of RNA sequences through self-supervised pretraining, with the goal of capturing transcript-level regulatory information. Early methods relied on handcrafted k-mer features or shallow encoders, which lacked the capacity to model long-range dependencies and complex structural signals\cite{alipanahi2015predicting}. To address these limitations, RNA-specific language models such as RNA-BERT \cite{akiyama2022informative} and RNA-FM \cite{chen2022interpretable} were introduced, trained on large-scale transcriptomic datasets to produce contextualized RNA embeddings. These models significantly improved the modeling of splicing signals, secondary structure contexts, and untranslated region motifs.
In parallel, general-purpose contextual learning frameworks such as data2vec \cite{baevski2022data2vec} have been adapted to RNA domains, offering a unified modality-agnostic representation space. Building on this direction, mRNA2vec \cite{zhang2025mrna2vec} proposes a joint encoding strategy for 5'UTRs and coding regions by integrating both contextual and structural cues, enabling holistic modeling of mRNA molecules. Although genomic models like DNABERT have shown strong performance in DNA sequence tasks, their architecture and training focus limit their applicability to RNA-specific regulatory interpretation. Recent multimodal extensions incorporating structural or epigenomic data further expand the landscape of transferable RNA representations.
\subsection{Supervised Deep Learning for 5'UTRs-Based Translation Modeling}
In contrast to general-purpose representation learning, another line of work has focused on directly modeling translation efficiency from 5'UTRs sequences using supervised deep learning. The 5'UTRs play a central role in post-transcriptional regulation, modulating translation initiation, ribosome recruitment, and start-site selection \cite{leppek2018functional}. Optimus \cite{sample2019human} leverages global-pooling convolutional networks (CNN) trained on mutagenesis data to predict MRL from fixed-length sequences, capturing the functional impact of sequence variants. FramePool \cite{karollus2021predicting} extends this approach with a length-agnostic CNN architecture that enables frame-wise ribosome occupancy prediction, offering fine-resolution insights into initiation dynamics.
To improve generalization across species and experimental conditions, MTrans \cite{zheng2023discovery} applies a multi-task framework with shared CNN encoders and task-specific GRUs. UTR-LM \cite{chu20245} , in turn, pretrains a Transformer-based model on large-scale collections of 5'UTRs sequences, learning transferable contextual embeddings for downstream translation-related tasks. Together, these models provide a strong foundation for 5'UTRs-based translation modeling, demonstrating reliable performance across a range of predictive tasks.
However, existing methods still face challenges in capturing interpretable, motif-aware dependencies and in modeling variable-length inputs without sacrificing biological relevance. These limitations motivate the development of specialized architectures that are both structurally adaptive and biologically transparent.

\begin{figure*}[t]
  \centering
  \includegraphics[width=0.95\linewidth]{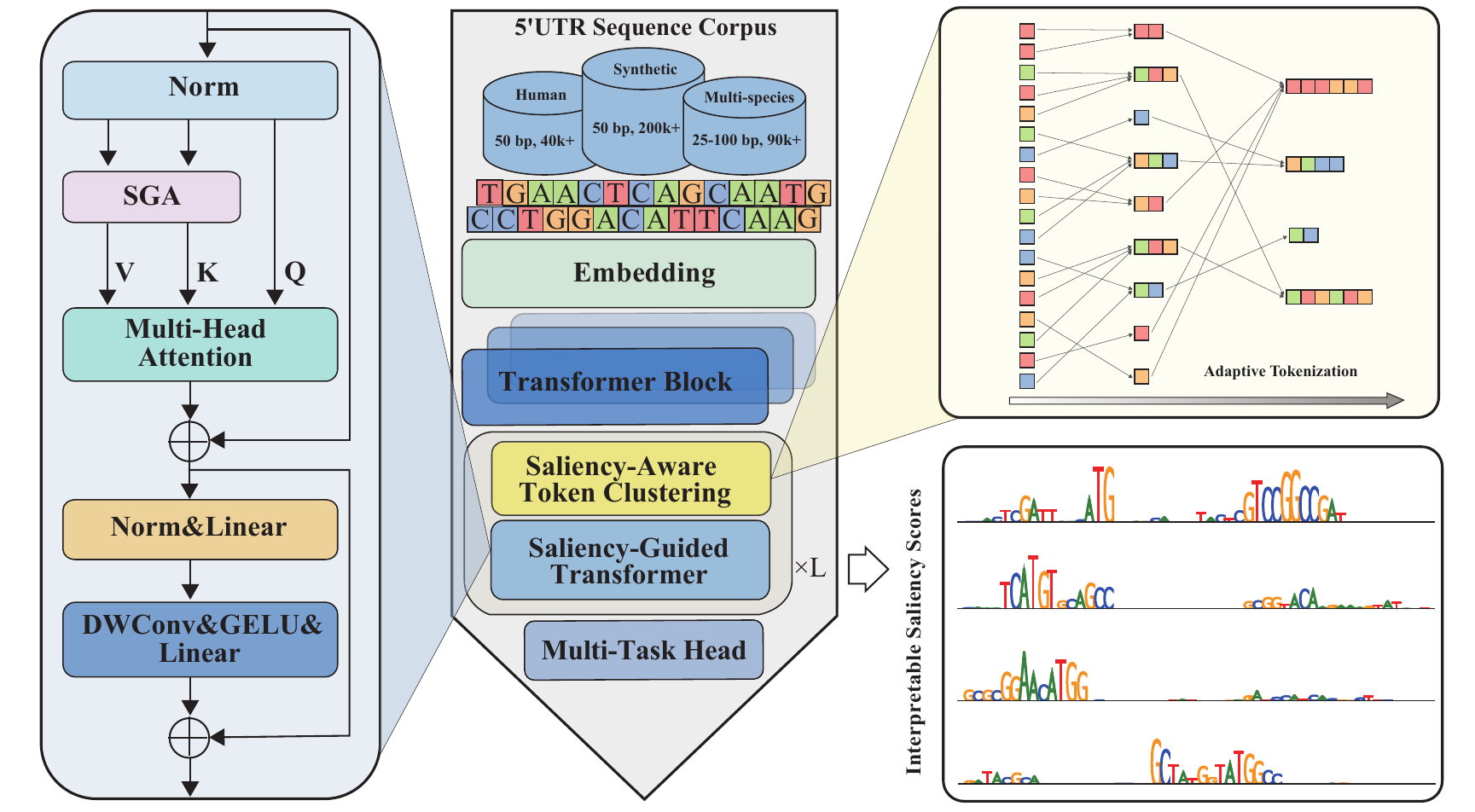}
\caption{Overview of UTR-STCNet. The model takes variable-length 5'UTRs sequences as input to predict mean ribosome load (MRL). It combines a Saliency-Aware Token Clustering (SATC) module for selecting informative tokens with a Saliency-Guided Transformer (SGT) block for modeling regulatory interactions. The output representations are biologically interpretable, enabling recovery of known sequence elements.}

  \label{fig:utr_stcnet}
\end{figure*}

\section{Method}
We present the architecture of UTR-STCNet in Figure.\ref{fig:utr_stcnet}. The model incorporates a basic Transformer block \cite{vaswani2017attention}, a Saliency Awared Token Clustering (SATC) Module, a Saliency Guided Transformer (SGT) Block, and multi-task prediction heads.
SATC module hierarchically merges relevant tokens into saliency-aware representatives, enhancing structural compactness and semantic coherence. These refined tokens are subsequently processed by the SGT, which compresses the representations of 5'UTRs into more compact, saliency-weighted embeddings, thereby improving computational efficiency and reducing model complexity while preserving the ability to extract critical regulatory features and capture long-range dependencies. 
We adopt a multi-task learning framework where the feature extraction backbone is shared across all tasks. To capture task-specific characteristics, we design independent prediction heads for each task. Each task-specific head consists of a lightweight feedforward network that maps the shared representation to the task-specific output space.
The task-specific heads then predict mean ribosome load based on the aggregated and interpretable representations.

\subsection{Problem Formulation}
We study the problem of predicting translation-associated activity from genomic sequences. Given a DNA sequence of length $L$, denoted as $\mathbf{s} = (s_1, s_2, \ldots, s_L)$ where each $s_i \in \{\mathrm{A}, \mathrm{C}, \mathrm{G}, \mathrm{T}\}$, our goal is to learn a mapping function $f_\theta: \mathcal{S} \rightarrow \mathbb{R}_{\ge 0}$ where \( \mathcal{S} \) denotes the space of all possible DNA sequences of length \( L \), that estimates the MRL. The prediction is denoted as $\hat{y}_{\mathrm{MRL}} = f_\theta(\mathbf{s})$.

To ensure consistent input representation, each sequence $\mathbf{s}$ is embedded into a matrix $\mathbf{X} \in \mathbb{R}^{L \times d}$, where $d$ is the dimensionality of the token embeddings based on nucleotides. 
The labeled training dataset is denoted as $\{(\mathbf{s}^{(i)}, y_{\mathrm{MRL}}^{(i)})\}_{i=1}^{N}$, where $N$ is the number of training samples.

\begin{table*}[htbp]
\caption{Performance Comparison of Single-Task Models Across Three 5'UTRs Translation Datasets }

\centering
\small  
\begin{tabular}{lccccccccc}
\toprule
\multirow{2}{*}{Model}
& \multicolumn{3}{c}{MPRA-U}
& \multicolumn{3}{c}{MPRA-H}
& \multicolumn{3}{c}{MPRA-V} \\
\cmidrule(lr){2-4} \cmidrule(lr){5-7} \cmidrule(lr){8-10}
& R\textsuperscript{\rm 2}$\uparrow$ & Spearman R$\uparrow$ & RMSE$\downarrow$
& R\textsuperscript{\rm 2}$\uparrow$ & Spearman R$\uparrow$ & RMSE$\downarrow$
& R\textsuperscript{\rm 2}$\uparrow$ & Spearman R$\uparrow$ & RMSE$\downarrow$ \\
\midrule

RNA-FM                   & 0.679    & 0.827    & 1.079    & 0.550    & 0.751    & 0.892   & 0.369    & 0.586    & 1.117  \\
OptimusN                  & 0.915  & 0.939 & 0.782 & 0.783  & 0.866 & 0.565 & 0.814  & 0.888 & 0.605 \\
Framepooling              & 0.895  & 0.909 & 0.818 & 0.796  & 0.881 & 0.543& 0.802  & 0.853 & 0.681 \\
MTrans                   & 0.941   & 0.954 & 0.447 & 0.812  & 0.902 & 0.526 & 0.845  & 0.909 & 0.552 \\
UTR-LM                   & 0.947    & 0.962    & 0.438    & 0.825    & 0.906    & 0.502  & 0.866    & 0.912    & 0.537  \\
mRNA2vec                   & 0.923    & 0.943    & 0.512    & 0.810  & 0.889 & 0.517    & 0.831    & 0.895    & 0.545\\
\textbf{UTR-STCNet}  &\textbf{0.956}  & \textbf{0.967} & \textbf{0.388} & \textbf{0.849}  & \textbf{0.914} & \textbf{0.471} & \textbf{0.907}  & \textbf{0.937} & \textbf{0.429} \\
\bottomrule
\end{tabular}%

\label{tab:single-task}
\end{table*}

\subsection{Saliency-Aware Token Clustering Module}

The adaptive tokenization process, implemented by the Saliency-Aware Token Clustering (SATC) module and illustrated in Figure.\ref{fig:utr_stcnet}, aggregates discrete input tokens into compact saliency-guided representations through two primary steps: (1) Saliency Gating, which identifies biologically informative tokens by assigning interpretable saliency scores and leverages these scores to guide subsequent clustering based on a density-distance potential; (2) Saliency-Preserving Centroid Synthesis, which aggregates the selected tokens within each cluster into unified representations via saliency-weighted averaging. This transformation effectively reduces token redundancy while preserving biologically informative features, resulting in more compact and interpretable embeddings that facilitate efficient modeling in subsequent layers.

\paragraph{Saliency Gating}
Before any compression, we apply saliency gating to decide which tokens are worth keeping.
Given a token vector $t_i\!\in\!\mathbb{R}^{d}$, and a learned weight vector $\mathbf{w}\!\in\!\mathbb{R}^{d}$, a single-head projection produces

\begin{equation}
  \alpha_i = \exp(\mathbf{w}^{\!\top} t_i) . \label{eq:saliency}
\end{equation}

In practice, $\alpha_i$ acts as a interpretable saliency scores, indicating the relative importance of token $i$.  
These weights are later used as positional importance factors for biological motifs (e.g., Kozak consensus, uAUG upstream start codons), offering interpretability without additional heads.

With the interpretable saliency scores fixed, we measure token cohesion and separation via a density-distance potential.
We first build a scale-normalised distance matrix
\begin{equation}
D_{ij} = \frac{ \| t_i - t_j \|_2 }{ \sigma \sqrt{d} },
\label{eq:distance}
\end{equation}
where \( t_i \in \mathbb{R}^d \) denotes the \( i \)-th token embedding and \( d \) is the feature dimension. 
The distance is normalized using the standard deviation of each embedding dimension, 
which reduces the dominance of high-variance features and improves the stability of distance computation.
We then estimate the local density \( \rho_i \) of each token using its \( k \)-nearest neighbors\cite{cover1967nearest}:
\begin{equation}
  \rho_i=\exp\!\Bigl(-\tfrac1k\sum_{j\in\mathcal{N}_k(i)} D_{ij}^{2}\Bigr),           \label{eq:density}
\end{equation}
where $\mathcal{N}_k(i)$ denotes the set of the $k$ nearest neighbors of token $i$. 
Then, for each token \( i \), we explicitly compute a distance indicator \( \delta_i \) representing the local density gap \cite{du2016study} as:
\begin{equation}
\delta_i =
\begin{cases}
\min_{j:\rho_j > \rho_i} D_{ij}, & \text{if } \exists j : \rho_j > \rho_i \\
\max_{j} D_{ij}, & \text{otherwise}
\end{cases}
\label{eq:sita}
\end{equation}
where \( \delta_i \) captures the minimal distance from token \( i \) to any other token with a higher local density. For the token with the highest local density, \( \delta_i \) is set to the maximum distance to any other token.
\begin{equation}
\quad \gamma_i = \rho_i \, \delta_i, \label{eq:gamma}
\end{equation}
where $\gamma_i$ represents the overall density--distance potential of token $i$ and guides the selection of representative centroids during clustering.

\paragraph{Saliency-Preserving Centroid Synthesis} 
After computing the unified density-aware score $\gamma_i$, we select the top $\lceil \tau N \rceil$ tokens to serve as centroids, where $\tau \in (0,1)$ is a tunable hyperparameter denoting the proportion of selected tokens out of the total number $N$. The selection is based directly on the scores $\gamma_i$, which reflect each token's contextual importance.
This efficient, single-pass procedure identifies representative regions without requiring iterative refinement steps such as Expectation–Maximization (EM) or clustering post-processing.

Once clustering assignments are fixed, tokens within each cluster $c$ are aggregated via a saliency-weighted average:
\begin{equation}
z_c = \sum_{i \in c} \left( \frac{\alpha_i}{\sum_{j \in c} \alpha_j} \right) t_i,
\label{eq:fusion}
\end{equation}
where $t_i $ is the contextual representation of token $i$ before clustering and $\alpha_i$ denotes the saliency scores obtained from the earlier Saliency Gating module (Eq.~\ref{eq:saliency}), explicitly capturing each token's contextual significance.

By leveraging saliency as a soft weighting in the fusion step, the SATC module selectively emphasizes tokens with high functional or semantic relevance, while down-weighting less informative ones, resulting in more compact and discriminative representations.

\subsection{Saliency Guided Transformer Block}
The Saliency-Guided Transformer (SGT) block refines token representations by emphasizing biologically informative features and reducing redundancy. It models local and long-range dependencies through Saliency-Guided Aggregation, which compresses token features using saliency scores, preserving crucial contextual and structural information. Additionally, it integrates Motif-Level Contextualization via depth-wise convolutions that efficiently capture position-specific motifs critical for regulatory elements governing gene expression.
\paragraph{Saliency Guided Aggregation}
The quadratic complexity of Transformer models poses a computational bottleneck for modeling long sequences, especially in 5'UTRs, where repetitive tokens often obscure essential regulatory information. To address this, we propose a Saliency-guided Aggregation (SGA) module, which reduces sequence length by aggregating Key and Value features according to their token-level saliency scores, thereby preserving structural and contextual information. 

Specifically, given Query, Key, and Value matrices $\mathbf{Q}, \mathbf{K}, \mathbf{V} \in \mathbb{R}^{B \times N \times C}$ and the saliency score matrix $\mathbf{P} \in \mathbb{R}^{B \times N \times 1}$ computed by the SATC module, we directly utilize these saliency scores as confidence weights to guide the subsequent local aggregation process.

For local aggregation, we partition the original sequences into segments $\Gamma_\ell$, each containing $r$ tokens, resulting in a reduced-length sequence with length $M = \lceil N / r \rceil$. Within each segment $\Gamma_\ell$, we denote the token-level Key and Value vectors at position $t$ as $\mathbf{k}_t, \mathbf{v}_t \in \mathbb{R}^C$, and the scalar saliency score $p_t \in \mathbb{R}$, extracted from the saliency matrix $\mathbf{P}$. We then perform saliency-weighted aggregation as follows:



\begin{figure}[htbp]
  \centering
  \includegraphics[scale=0.56]{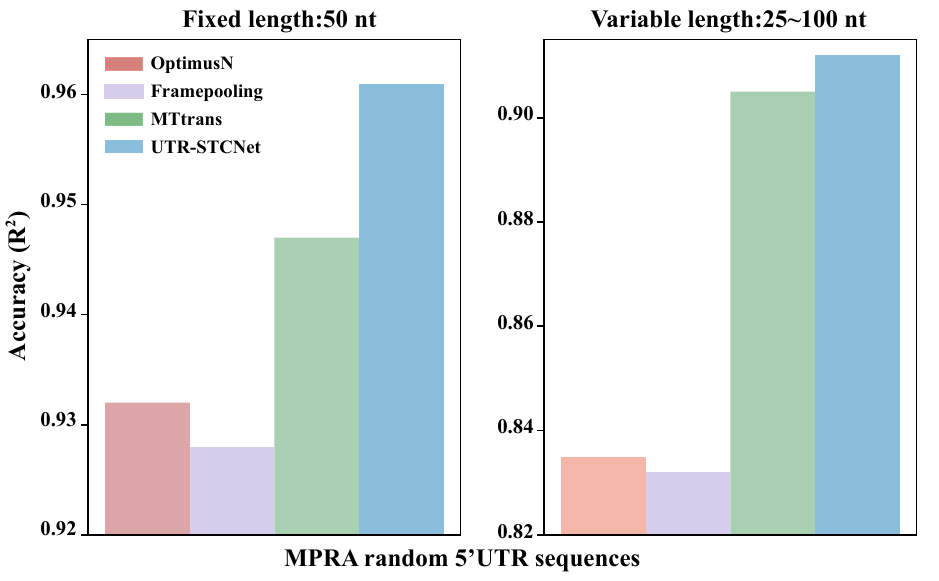}
  \caption{Prediction performance of UTR-STCNet on MRL. Evaluation was conducted on MPRA datasets with fixed-length sequences (left, n = 20,000) and variable-length sequences (right, n = 7,600). Performance was measured using the coefficient of determination ($\text{R}^2$) between observed and predicted MRL values on held-out test sets. }
  \label{fig:multi-task}
\end{figure}

\begin{equation}
    \begin{split}
    & \widetilde{\mathbf{K}}_\ell = 
     \frac{\sum_{t \in \Gamma_\ell} p_t \odot k_t}{\sum_{t \in  \Gamma_\ell} p_t + \varepsilon},\\
    &\widetilde{\mathbf{V}}_\ell =
    \frac{\sum_{t \in \Gamma_\ell} p_t \odot v_t}{\sum_{t \in \Gamma_\ell} p_t + \varepsilon},\\
    \end{split}
    \label{eq:casr_pooling}
\end{equation}



where $\varepsilon > 0$ ensures numerical stability, and $\odot$ denotes element-wise multiplication. Stacking the aggregated vectors across all segments yields the reduced-length Key and Value matrices $\tilde{\mathbf{K}}, \tilde{\mathbf{V}} \in \mathbb{R}^{B \times M \times C}$, preserving salient structural dependencies. Additionally, this method substantially reduces the spatial dimension of Key and Value, leading to decreased computational and memory overhead, and effectively models local continuity, thereby better preserving local token correlations.

In the subsequent attention layer, the clustered token representations serve as the Query $\mathbf{Q}$, while the compressed Key and Value representations ($\widetilde{\mathbf{K}}, \widetilde{\mathbf{V}}$) are used as the context.
To ensure that critical tokens maintain strong influence, we directly integrate token-level saliency scores into the attention logits by constructing an attention bias matrix $\mathbf{P}_{\mathrm{attn}}$. Then, the attention is computed as:
\begin{equation}
    \text{Attention}(\mathbf{Q}, \widetilde{\mathbf{K}}, \widetilde{\mathbf{V}}) =
    \text{softmax} \left( \frac{\mathbf{Q} \widetilde{\mathbf{K}}^\top}{\sqrt{d_k}} + \mathbf{P}_{\mathrm{attn}} \right) \widetilde{\mathbf{V}}.
    \label{eq:casr_attn}
\end{equation}
This formulation enables high-saliency tokens to retain dominance in downstream attention, even after SGA, thereby achieving a balance between compression efficiency and information retention.

\begin{table}[h]
\caption{Effectiveness of different modules in UTR-STCNet.}
\centering
\begin{tabular}{lccc}
\toprule
\multirow{2}{*}{Model} & \multicolumn{3}{c}{MPRA-H} \\
\cmidrule(lr){2-4}
& R\textsuperscript{\rm 2}$\uparrow$ & Spearman R$\uparrow$ & RMSE$\downarrow$  \\

\midrule
w/o STCM \& SGA         & 0.826 & 0.889 & 0.533 \\
w/o SGA  & 0.837 & 0.905 & 0.486  \\
w/o STCM        & 0.833 & 0.896 & 0.504  \\
\specialrule{0.03em}{0pt}{2pt}  
UTR-STCNet         & \textbf{0.849} & \textbf{0.914} & \textbf{0.471}  \\

\bottomrule
\end{tabular}
\label{tab:ablation}
\end{table}

\paragraph{Motif-Level Contextualization}
Although the aforementioned SATC module compresses the sequence along its sequence dimension, the feed-forward layer in standard transformers remains strictly position-wise and therefore struggles to capture short-range compositional patterns such as 3 to 5 nucleotide motifs, which are critical for 5'UTRs regulation. To address this, we redesign the feed-forward sub-layer by incorporating a depth-wise convolutional structure \cite{chollet2017xception} that effectively captures both known and novel motif signals, as well as other functionally relevant sequence elements. Specifically, the revised module applies a 1D depth-wise convolution followed by a GELU activation\cite{hendrycks2016gaussian}, with a residual connection to ensure gradient stability.
Depth-wise convolutions are particularly well-suited for modeling local dependencies in sequence data, as they apply channel-specific filters to capture fine-grained patterns with high parameter efficiency. The projection weights before and after the convolution are shared across tokens, preserving computational efficiency while enabling each channel to extract distinct local motif features.

This depth-wise convolutional mechanism introduces positional awareness at the motif level in a channel-efficient manner, complementing self-attention without requiring additional positional encodings. This mechanism further enables adaptation to varying input channel sizes following token aggregation, while preserving the spatial arrangement of tokens and capturing multi-scale features within the sequence. It integrates seamlessly with the channel-based token representations produced by the preceding SGA, enabling biologically meaningful context modeling into the SGT backbone. This enables the capture of functional differences across genes and complex regulatory contexts, enhances the modeling of dependencies between regulatory elements and sequence features, and ultimately improves the interpretability of functional signals in RNA sequences.

\section{Experiments}
\subsection{Dataset}
\begin{figure}[htbp]
  \centering
  \includegraphics[scale=0.67]{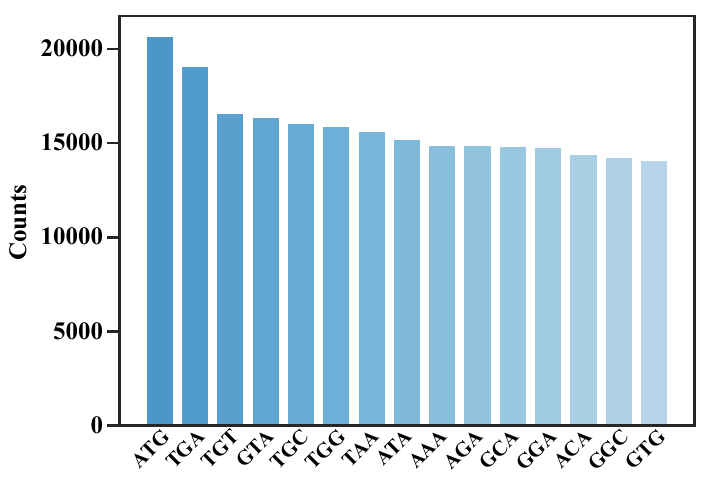}
  \caption{Top enriched 3-mer motifs identified within high-saliency regions. High-saliency regions were first identified as stretches of consecutive nucleotides with elevated saliency scores. Within these regions, a 3-mer sliding window (step size = 3) was applied to extract all possible 3-mer motifs. A 3-mer was counted once per sequence if it appeared in a high-saliency region and exceeded a predefined score threshold. The top 15 3-mers with the highest frequency across all sequences are shown.}
  \label{fig:Fig3}
\end{figure}
We evaluated the proposed UTR-STCNet model using three publicly available 5'UTRs translation datasets: MPRA-U, MPRA-H, and MPRA-V. These datasets were all derived from massively parallel reporter assays (MPRAs)\cite{cuperus2017deep}, providing abundant sequence information and corresponding expression levels to support model training and validation. 
We first ranked all sequences based on total read counts obtained from sequencing, and selected the top 20,000 most highly expressed sequences from the MPRA-U dataset as test samples. Similarly, the top 7,600 highly expressed human-derived sequences from the MPRA-H dataset were selected as the test set. 
For the MPRA-V dataset, due to the variability in sequence lengths, we adopted a stratified random sampling strategy: sequences were divided into length bins ranging from 25 to 100 nucleotides, and from each bin, 100 randomly generated sequences and 30 natural human sequences were selected, resulting in a total of 9,880 test sequences. 
The remaining unselected sequences from each dataset were used to construct the combined training and validation sets.

\subsection{Baselines}
We compare UTR-STCNet with the following seven state-of-the-art baseline models:

\begin{itemize}

    \item \textbf{Optimus} \cite{sample2019human}  employs convolutional networks with residual blocks to forecast mean ribosome load across fixed-length 5' UTRs.
    \item \textbf{FramePool} \cite{karollus2021predicting} is a length-agnostic CNN baseline that predicts 5'UTRs mean ribosome load using frame pooling.
    \item \textbf{RNA-FM}\cite{chen2022interpretable} is a foundation model that captures RNA structure, 3D proximity, and RBP interactions from unannotated sequences.
    \item \textbf{MTtrans} \cite{zheng2023discovery} shares a convolutional encoder across tasks while task‑specific GRU towers yield length‑agnostic, cross‑assay translation‑rate predictions. 
    \item \textbf{UTR-LM} \cite{chu20245} is a transformer-based language model that enables unified functional modeling and pretraining over 5'UTRs sequences.
    \item \textbf{mRNA2vec}(Zhang et al., 2025) is a general-purpose pretrained model based on data2vec jointly encoding 5'UTRs and CDS with contextual and structural signals for mRNA representation.

\end{itemize}

\begin{figure}[htbp]
  \centering
  \includegraphics[scale=0.36]{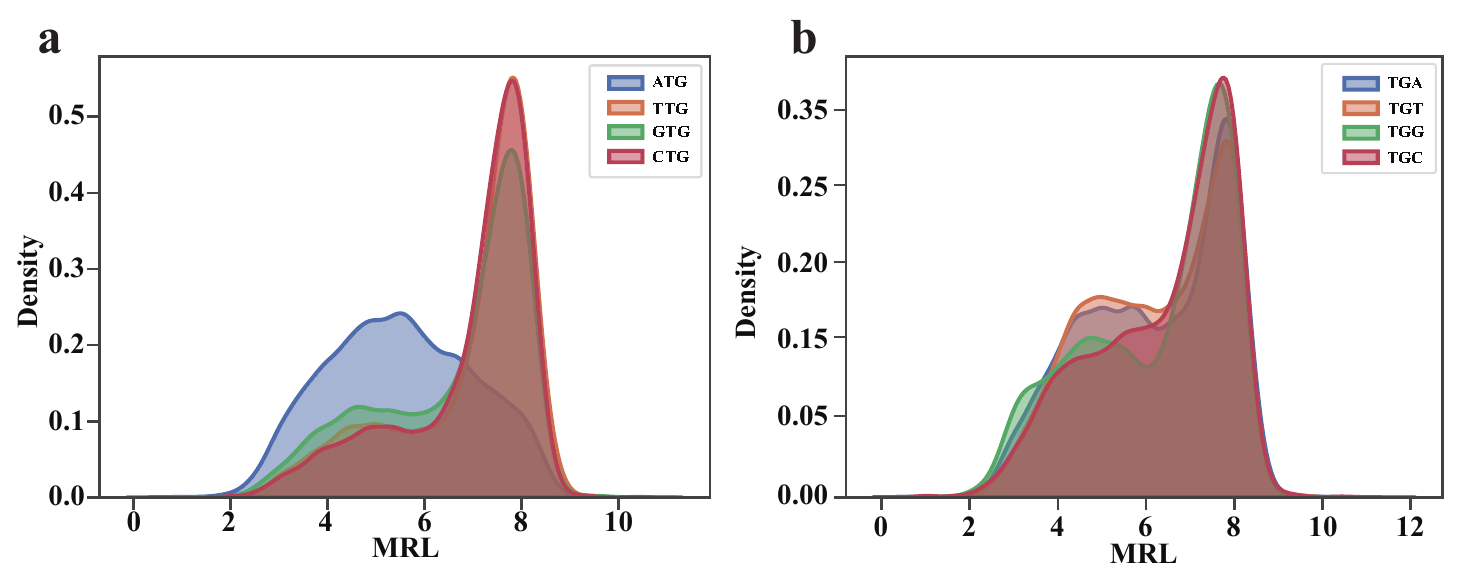}
  \caption{Distribution of saliency scores across TG-related groups. Density plots show the distribution of token-level saliency scores for different TG-related groups in panels a and b.}
  \label{fig:Fig4}
\end{figure}

\subsection{Performance Comparison}
\paragraph{Single-Task Evaluation}
Table~\ref{tab:single-task} demonstrates that UTR-STCNet consistently achieves state-of-the-art performance across the three 5'UTRs translation datasets: MPRA-U, MPRA-H, and MPRA-V. The model is comprehensively evaluated in terms of its generalization capability and predictive accuracy.
On the MPRA-U dataset, UTR-STCNet achieves approximately 11.4\% lower prediction error than UTR-LM, reflecting superior modeling accuracy.This result suggests that the model more effectively captures the complex nonlinear relationships between sequence features and expression levels, maintaining superior predictive performance even for sequences with high expression or strong regulatory signals.
On the MPRA-H dataset, UTR-STCNet outperforms mRNA2vec by around 4.8\%, highlighting its enhanced ability to model gene expression on real human-derived sequences. This performance gain demonstrates that UTR-STCNet is better equipped to learn intrinsic regulatory features and expression patterns in biologically authentic contexts, providing improved reliability and application potential in human transcriptome modeling.
On the MPRA-V dataset, UTR-STCNet achieves a 3.1\% improvement in Spearman correlation compared to MTrans. This dataset includes variable-length sequences, designed to rigorously evaluate the model’s robustness to input heterogeneity and its ability to capture long-range dependencies. The observed performance further validates UTR-STCNet’s architectural advantages in modeling complex contextual structures and its practical applicability to diverse 5'UTRs length distributions found in real biological scenarios.

\paragraph{Multi-Task Evaluation}
As shown in Figure~\ref{fig:multi-task}, UTR-STCNet consistently achieves superior performance in multi-task settings on both fixed-length and variable-length synthetic MPRA datasets. On the fixed-length dataset (50~nt), UTR-STCNet effectively integrates multi-task information and achieves an $R^2$ score of 0.965, significantly outperforming the multi-task baseline model MTrans, demonstrating its stronger capability in modeling gene expression.
On the more challenging variable-length dataset (25--100~nt), UTR-STCNet also outperforms all three baseline models, indicating its ability to dynamically adapt to input length variations and fully leverage contextual structures, thereby substantially enhancing prediction accuracy and modeling stability.
Overall, UTR-STCNet not only delivers superior performance in multi-task scenarios, but also demonstrates greater robustness and generalization in handling variable-length inputs, which remain a fundamental challenge in real biological sequences, thereby highlighting its promising potential for practical applications.

\begin{figure}[htbp]
  \centering
  \includegraphics[scale=0.45]{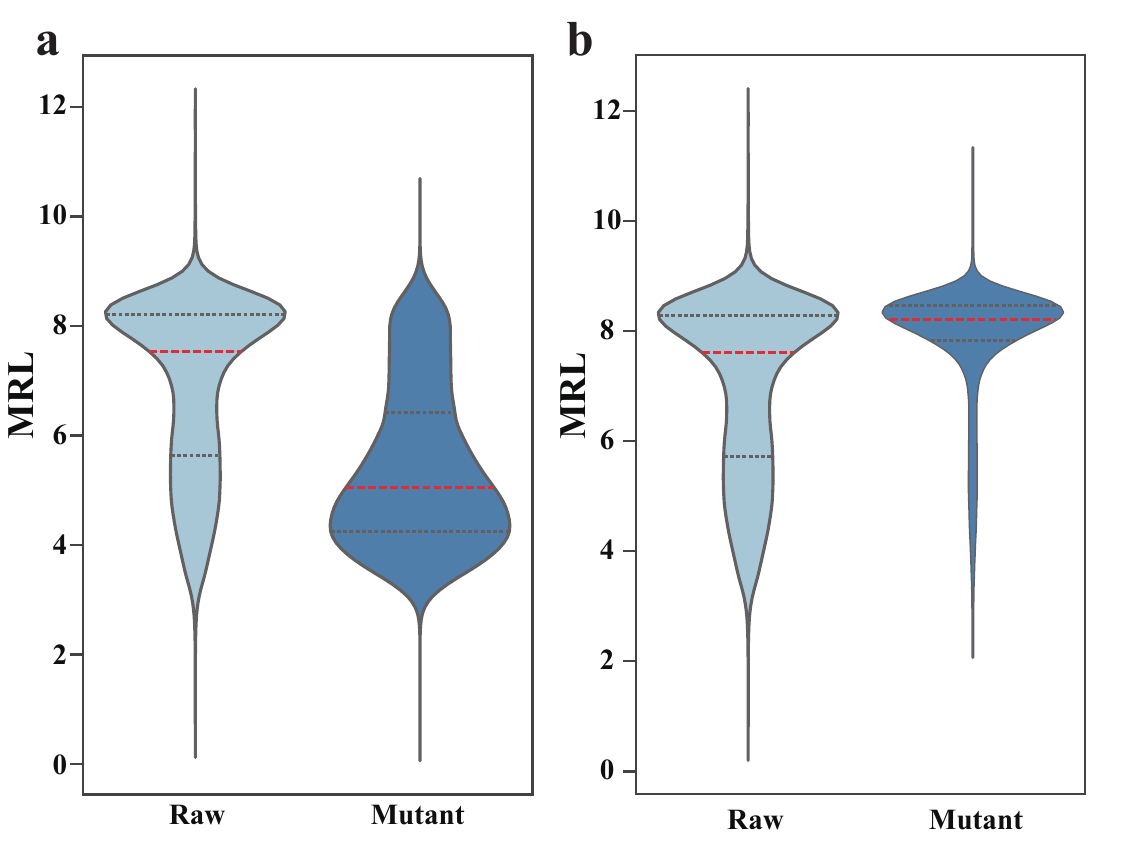}
  \caption{In silico mutagenesis reveals the functional impact of TG-related motifs on MRL predictions.
a, Predicted MRL values after replacing the nucleotide preceding the `TG' motif with adenine (A). Dark blue: mutated sequences; light blue: original experimental sequences.
b, Predicted MRL values after replacing the `TG' motif itself with non-TG variants. Dark blue: substituted sequences; light blue: original sequences.}
  \label{fig:Fig5}
\end{figure}

\paragraph{Ablation Study}
To evaluate the contribution of each core module in UTR-STCNet, we perform an ablation study on the MPRA-H dataset, as shown in Table~\ref{tab:ablation}. Specifically, we examine three ablation variants: removing both the STCM and the SGA, removing only SGA while retaining STCM, and removing only STCM while keeping SGA.
The results indicate that the combination of STCM and SGA yields the strongest performance, with an $R^2$ score of 0.849, Spearman correlation of 0.914, and RMSE of 0.471. In comparison, removing the SGA module alone reduces the $R^2$ to 0.837, showing that SGA plays a crucial role in preserving local continuity and guiding attention-weighted channel fusion. Similarly, removing STCM reduces the $R^2$ to 0.833, confirming that saliency-aware token clustering contributes to compressing redundant regions and enhancing structural coherence. When both modules are removed, the model performs worst, with $R^2$ dropping to 0.826 and RMSE increasing to 0.533. This highlights the limitations of directly applying vanilla Transformers to long UTRs sequences without guided compression or context-aware aggregation.
These findings validate the effectiveness of both modules. STCM improves input compactness through structural guidance, while SGA enhances downstream predictions by emphasizing important signals during aggregation. Together, they contribute synergistically to the model's overall accuracy and generalization.

\subsection{Biological Relevance of Model-Identified Regulatory Features}
\paragraph{Identification of High-Scoring Regions and Key Motifs}
\begin{figure}[htbp]
  \centering
  \includegraphics[scale=0.5]{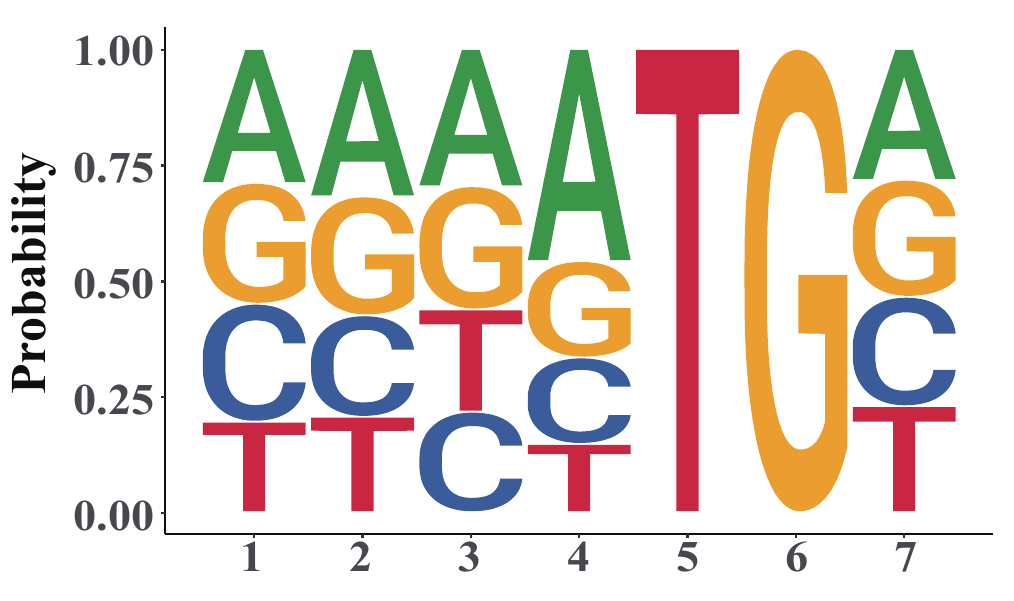}
  \caption{From the high-saliency regions, sequences centered on the dinucleotide `TG' were extracted by taking 4 nucleotides upstream and 1 nucleotide downstream. The nucleotide frequencies flanking `TG' were then computed to analyze base composition patterns surrounding this motif.}
  \label{fig:fig6}
\end{figure}
To investigate the regulatory features prioritized by UTR-STCNet, we designated high-scoring regions as contiguous stretches of three or more nucleotides with saliency scores exceeding 0.5, yielding a total of 22,393 candidate motifs. To systematically characterize local sequence patterns within these regions, we conducted a sliding window analysis using a 3-mer window. We computed the frequency of each 3-mer across all high-scoring regions and found that sequences containing the dinucleotide `TG' were markedly overrepresented Figure \ref{fig:Fig3}, suggesting that `TG'-containing motifs may serve as important signals for MRL prediction. To validate this finding, we repeated the analysis using a 2-mer window. Consistent with the 3-mer results, `TG' remained the most enriched 2-mer, reinforcing its role as a recurrent sequence feature learned by the model.

\subsection{Impact of Adjacent Nucleotides to TG on MRL Prediction.}
\paragraph{MRL Distributions of TG-Associated Nucleotide Combinations.} We analyzed the distribution of experimentally measured MRL values for different TG-associated nucleotide combinations. As shown in Figure \ref{fig:Fig4}a, sequences containing the ATG trinucleotide exhibited significantly lower MRL values compared to other combinations, suggesting that ATG may act as a translation-suppressive element \cite{andreev2022non}. In contrast, varying the nucleotide immediately following TG (A, C, G, or T) showed no substantial differences in MRL distributions Figure \ref{fig:Fig4}b, indicating that among trinucleotides containing TG, the presence of upstream ATG (uATG) is specifically associated with a negative regulatory effect \cite{tsotakos2015regulation}.

\indent To validate the negative regulatory effect of ATG on MRL, we conducted an extreme test by replacing any non-A base immediately preceding each TG with A, while leaving other bases unchanged. This resulted in a significant overall decrease in MRL Figure \ref{fig:Fig5}a, further supporting the conclusion that ATG negatively impacts translation efficiency. Finally, to confirm whether the inhibitory effect is caused by the ATG triplet or the TG sequence itself, we randomly replaced all TGs with non-TG combinations Figure \ref{fig:Fig5}b. The prediction results indicated that while TG slightly reduced reaction efficiency, the primary inhibitory effect was attributed to the ATG sequence.

\paragraph{Feature Extraction Near Translation Initiation Sites.}
To evaluate whether the model can capture regulatory patterns near translation initiation sites, we analyzed the nucleotide composition surrounding high-saliency TG motifs. For each TG site within high-scoring regions, we extracted four upstream and one downstream nucleotide and computed the positional nucleotide frequency distribution. The analysis revealed that the surrounding sequence context closely resembles the classical Kozak motif Figure \ref{fig:fig6} \cite{kozak1986point}, indicating that the model effectively captures known regulatory features associated with translation efficiency.

\section{Conclusion}
In this work, we propose UTR-STCNet, a deep learning framework for modeling translational regulation from 5'UTRs sequences. By combining a Saliency-Aware Token Clustering module with a Saliency-Guided Transformer block, UTR-STCNet captures long-range dependencies and key regulatory motifs with high interpretability. Extensive experiments demonstrate that the model consistently outperforms existing approaches across species and cell types, achieving state-of-the-art performance in MRL prediction. In addition to accurate inference, UTR-STCNet reveals biologically meaningful patterns, recovering canonical elements such as uAUGs and Kozak motifs. These results highlight UTR-STCNet as a promising tool for advancing mRNA design and understanding translation control.

\section*{Acknowledgment}
We are grateful to the members of the Yu Laboratory and many colleagues for their valuable comments and suggestions. This work was supported by the Major Project of Guangzhou National Laboratory (Grant No.GZNL2024A01003), the National Natural Science Foundation of China (Grant No.32470634), and the Guangdong Basic and Applied Basic Research Foundation (Grant No.2024B1515020080).

\bibliographystyle{IEEEtran.bst}
\bibliography{UTR.bib}

\end{document}